\documentclass[a4paper,10pt,twoside]{cpc-hepnp}

\usepackage{multicol}
\usepackage{graphicx}
\usepackage{booktabs}
\usepackage{amssymb,bm,mathrsfs,bbm,amscd}
\usepackage[tbtags]{amsmath}
\usepackage{lastpage}

\begin{document}
\fancyhead[co]{\footnotesize ...
{ }} \footnotetext[0]{Submitted to Chinese Physics C }

\title{Strange Quarkonium States at BESIII\thanks{ Supported in part
by the CAS/SAFEA International Partnership Program for Creative
Research Teams, CAS and IHEP grants for the Thousand/Hundred Talent programs and National Natural Science Foundation of China under Contracts No. 11175189.}}

\author{%
\quad LIU Pei-Lian$^{1;1)}$\email{liupl@ihep.ac.cn}%
\quad FANG Shuang-Shi$^{1;2)}$ \email{fangss@ihep.ac.cn}%
\quad LOU Xin-Chou$^{1,2}$ 
}
\maketitle

%

\address{%
$^1$ Institute of High Energy Physics, Chinese Academy of
Sciences, Beijing 100049, China\\
$^2$ University of Texas at Dallas, Richardson, Texas 75080-3021, USA\\
}

\begin{abstract}
We present physics opportunities and topics with the $s\bar{s}$ states
(strangeonia) that can be studied with the BESIII detector operating at the BEPCII collider.
Though the $\phi$ and $\eta/\eta^\prime$ states have long been established experimentally,
only a handful of strangeonia are well known,
in contrast to the rich $c\bar{c}$ charmoium system.
An overview of the $s\bar{s}$ states and their experimental status is presented in this paper.
The BESIII experiment has collected the world's
largest samples of $J/\psi$, $\psi(2S)$, $\psi(3770)$, and direct
$e^+e^-$ annihilations at energies below the $J/\psi$ and above 3.8\,GeV, and will continue to accumulate high quality, large integrated luminosity in the $\tau$-charm energy region. These data, combined with the excellent performance of the BESIII detector, will offer unprecedent opportunities to explore the $s\bar{s}$ system. In this paper we describe the experimental techniques
to explore strangeonia with the BESIII detector.
\end{abstract}



\begin{multicols}{2}

\section{Introduction}
Half a century ago the quark model was introduced to describe the
large array of hadrons, within which mesons are composed
of $q\bar{q}$ bound together by the strong interaction and
baryons consist of three quarks. Since then the quark model has
offered a useful tool in understanding the substructures of hadrons
and led to the advent of Quantum Chromodynamics (QCD) for
describing the strong interaction. With the development of
accelerators and detectors, hundreds of hadrons have been discovered and
the quark model provides a good description of the
observed hadrons, in particular the ground states and the heavy
quarkonium states. Taking charmonium spectroscopy, bound
states of the charmed quark and antiquark ($c\bar{c}$), as an example,
it reflects a beautiful regularity and numerous
electromagnetic and strong transitions. But there are
still many unseen states predicted by the quark model, 
in particular the $s\bar{s}$ states, for which the experimental
situation is not well understood.

Being the bound state of a light $s$ and anti-$s$ quark, a strangeonium state
is related to long-range (i.e., confinement) interactions,
which provide information on non-perturbative QCD in the low energy region,
where the heavy quark effective theory is not applicable.

The study of the strangeonium mesons are of particular interest since they
are a bridge between the light $u$, $d$ quarks and the heavy $c$, $b$ quarks. In
addition, the strangeonium spectrum also helps identify the exotics (e.g.,
glueball, hybrid, multi-quark states) which can decay into the same final
states.
Within the framework of the relativistic quark model with QCD~\cite{r-q-model}, a spectroscopy similar to heavy quarkonia is expected for
strangeonium ($s\bar{s}$) states. Their decays were studied in detail with
the $^{3}P_{0}$ model~\cite{p3model}, which is a phenomelogical theory for
describing light meson decays.
The $s\bar{s}$ spectroscopy with spin $J<4$ predicted by the $^{3}P_{0}$ model
is shown in Fig.~\ref{strange}. Only seven states (underlined with the solid lines) have been
established and many members of the spectrum are still missing.

The poor experimental situation is largely due to the small fractions of $s\bar{s}$ states produced among hadrons,
and their large width. 
At present the experimental information on the
strangeonium states mainly came from the diffractive photoproduction reactions $\gamma p \rightarrow X
p$, strangeness exchange reactions $K^-p\rightarrow X \Lambda$, and $e^+e^-$ collisions.
Given our unsatisfactory knowledge of strangeonium, the search for these
missing states is critical for understanding the $q\bar{q}$
interaction.

\begin{center}
\includegraphics[width=9.0cm,height=8.5cm]{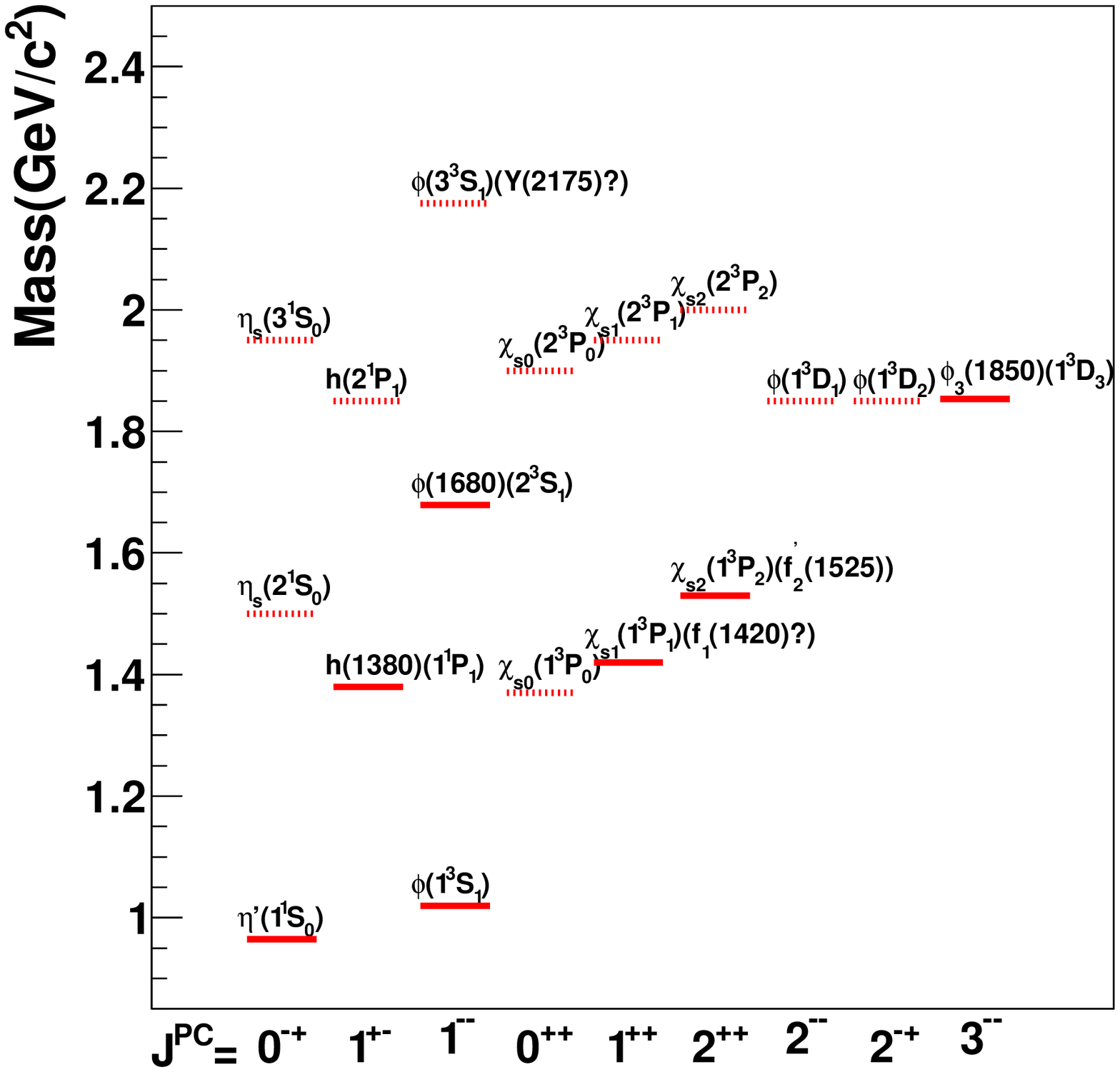}
\figcaption{\label{strange} The strangeonium family.
 }
\end{center}

In this paper, we will review the experimental status of
strangeonia, show the possible $J/\psi$ and
$\psi^\prime$ decays best suited for the search and study of new strangeonia, and propose approaches to search for new
strangeonium-like states similar to the charmonium-like states ($X$, $Y$, $Z$), which were
discovered at the Babar, Belle and BESIII experiments. 
We identify the $e^+e^- \rightarrow (s\bar{s})(s\bar{s})$ interaction as an especially important data sample to reconstruct new $s\bar{s}$ mesons with high efficiency in a mode independent way.

\section{Strangeonia status}

\subsection{$\eta^\prime$ ($1^1S_0$)}
The strangeonium spectrum is shown in
Fig.~\ref{strange}. Starting with the $1^1S_0$ state, the $\eta^\prime$
was discovered a half century ago in bubble chamber
experiments~\cite{etaprime}. Since then the $\eta^\prime$, being a
pure SU(3) singlet, has attracted both theoretical and
experimental attention due to its special role in understanding low
energy QCD. Because of the chiral anomaly,
the $\eta^\prime$ is a non-Goldstone boson which distinguishes itself
from the others in the ground pseudoscalar nonet. Therefore the
$\eta^\prime$ provides a unique field to test the fundamental
symmetries and the predictions for chiral perturbative theory. In
addition, the $\eta-\eta^\prime$ mixing issue remains. 
Of interest is the gluonic content in $\eta^\prime$, which makes this case more
complicated. Many interesting and important issues are extensively
discussed in Ref.~\cite{11024999}.

In general our present knowledge of the $\eta^\prime$ is based on
limited statistics; only its most dominant decays have been observed,
and are not well measured. To perform high statistics
measurements, the study of $\eta^\prime$ decays has already been listed
in the physics programs of many experiments, including CLAS, Crystal
Ball, WASA-at-COSY and KLOE-2. In addition to $\tau$-charm physics,
BESIII also has the capability to investigate $\eta^\prime$
decays via $J/\psi$ radiative or hadronic decays.
Based on a sample of $1.3\times 10^{9}$ $J/\psi$ events, BESIII has made
many important contributions to the study of $\eta^\prime$ decays,
including the measurements of $\eta^\prime$ hadronic
decays~\cite{besetaprime}, and the search for rare or forbidden
decays~\cite{besetaprime2}.

\subsection{$\phi$ ($1^{3}S_{1}$)}
Another well established strangeonium is the $\phi(1020)$, which is the
$1^{3}S_{1}$ state. The $\phi$ meson was first seen in a bubble
chamber experiment at Brookhaven in 1962, was subsequently determined to be a vector meson with $J^{PC}=1^{--}$, and could
be accommodated as an SU(3) singlet, supplementing an octet of other
vector mesons (e.g., $\rho$, $K^*$ and $\omega$). To explain that
$\phi$ prefers to decay into two kaons but not $\rho\pi$, the
Okubo-Zweig-Iizuka (OZI) rule~\cite{ozi} was introduced. The $\phi$
contains a $s\bar{s}$ pair; when it decays, the
strange quarks have to go somewhere, and the kaon pair route is the only
possibility. 
Since it has a quite narrow width and could be directly produced in $e^+e^-$
collisions, a $\phi$ factory, named DA$\Phi$NE, was built to study
its properties using the abundant production of $\phi$ collected with
the KLOE detector, which had made a significant contribution to
kaon and hadronic physics in the low QCD energy region. 
An upgrade of DA$\Phi$NE and the KLOE detector were performed.
The luminosity was designed to be improved by a factor of 3 and the
detector was also upgraded accordingly. A review of the physics
with the KLOE-2 detector at DA$\Phi$NE is given in
Ref.~\cite{kloe2}.


\subsection{ $h_1(1380)$ ($^1P_1$) }
The $^1P_1$ strangeonium state, $h_1(1380)$, is still a poorly known meson.
In 1988, the first evidence of the $^1P_1$ strangeonium state was seen by the LASS spectrometer at SLAC, from a Partial Wave Analysis (PWA) of $K^- p\rightarrow K^0_S K^{\pm} \pi^{\mp}\Lambda$~\cite{PLB201-573}.
It was confirmed in the $p\bar{p}\rightarrow K_L K_S\pi^0\pi^0$ by the crystal barrel detector at LEAR
with a mass of $M=(1440\pm60)$\,MeV/$c^2$ and a width of
$\Gamma=(170\pm80)$\,MeV~\cite{PLB415-280}. Due to the severe suppression by the phase
space, its dominant decay is $K^*K$. To date only these two experiments have
observed this state, and further confirmation from other experiment is strongly needed. 
There is some theoretical interest in $h_1(1380)$. In particular, Ref.~\cite{h1-theo} discusses the origin of this state in chiral dynamics from the interaction of the vector-pseudoscalar.

\subsection{ $f_1(1420)$ ($^3P_1$) }
The first evidence for the $f_1(1420)$ was seen in the $K^*\bar{K}$ mass spectrum in $\pi^- p$
reactions~\cite{NPB169-1}, but it is questionable because the structures around 1.4$-$1.5\,GeV
(e.g., $\eta(1405)$, $\eta(1475)$) are found to be complicated.
Detailed analyses~\cite{f1420-confirm} confirmed its existence and the spin-parity is determined to
be $J^{PC}=1^{++}$. The observations of $f_1(1420)$ in central production~\cite{f1420-central} provides
unambiguous evidence that $f_1(1420)$ has spin 1 and suggests a non-strange quark component in
$f_1(1420)$ despite its decay to $K^*K\pi$. 
In addition, the $f_1(1420)$ is not seen in the strange exchange $K^- p$ interactions, in which another meson, the $f_1(1510)$, possibly having large
$s\bar{s}$ components, is observed. Therefore, it has been suggested that the $f_1(1420)$  could be a
hybrid~\cite{hybrid}, a four quark~\cite{fourquark} or a molecular-like
state~\cite{molecul}.

This state has also been investigated in radiative and hadronic
decays of the $J/\psi$. MARK III~\cite{mark3-hadron} reported a structure around 1.42\,GeV observed in the $K^*\bar{K}$
mass spectrum in $J/\psi\rightarrow\omega K^*\bar{K}\pi$, but no similar peak was seen in
$J/\psi\rightarrow\phi K\bar{K}\pi$. These results were confirmed by the
BESII experiment~\cite{bes2-f1420} using the 58 million $J/\psi$ events. The mass and width obtained
from a fit to the $K^*K\pi$ mass are consistent with those of the $f_1(1420)$. The spin-parity is not determined here due to the large background.
The amplitude analyses of
$J/\psi\rightarrow \gamma K\bar{K}\pi$ indicates that the structure observed
around 1.42\,GeV in the $K\bar{K}\pi$ mass spectrum is a mixture of two or
three states~\cite{prl065-1990}.

The $f_1(1510)$ could be the $^3P_1$ $s\bar{s}$ state
instead of the $f_1(1420)$ due to its production in hadronic $K^-p$
experiments. However, the absence of the $f_1(1510)$ in hadronic
and radiative $J/\psi$ decays makes this case complicated. Given the
complexity in the $K\bar{K}\pi^0$ system, it is important to extract
these states in $J/\psi$ decays using high statistics data which will
enable PWA. In addition to $K^*K$, the observation of new decay modes, e.g., $\pi^+\pi^-\eta$ and $4\pi$, could also provide
valuable information on the $f_1(1420)$. Recently,
BESIII reported evidence for $f_1(1510)$ observed in
$J/\psi\rightarrow\gamma\pi^+\pi^-\eta^\prime$ decays~\cite{bes3}.

\subsection{ $f^\prime_2(1525)$ ($^3P_2$)}
The $f^\prime_2(1525)$ is widely accepted as the $^3P_2$ state, which was
first seen in $\pi^- p\rightarrow K^0_S K^0_S n$ collisions with limited
statistics. Since then this state has been observed in many different production processes, including
$K^-p$ collisions, $e^+e^-$ annihilations, $p\bar{p}$ annihilations, 
and $ep$ collisions. Given that it decays dominantly into a
$K\bar{K}$ instead of a $\pi\pi$ and is only observed in
$J/\psi\rightarrow\phi K\bar{K}$ decays, it seems to be a $s\bar{s}$ meson. 
A new idea that the $f^\prime_2(1525)$ is dynamically generated from the vector-vector interaction has been introduced in Ref.~\cite{1525-idea}. Support for this idea can be found in Ref.~\cite{1525-theo}, in which approaches to study the production of $f^\prime_2(1525)$ and other resonances ($f_{0}(1370)$, $f_{0}(1710)$, $f_{2}(1270)$, $K^*_{2}(1430)$) in the $J/\psi$, $\psi'$ and $\Upsilon(nS)$ decays are broadly discussed.

\subsection{$\phi(1680)$ ($2^3S_1$)}
The $\phi(1680)$ is well established in $e^+e^-$ production.
It was reported in $e^+e^-\rightarrow K_SK^\pm\pi^\mp$~\cite{dm1} and
subsequent analyses~\cite{dm1-2} found the small rate of $\phi(1680)$
coupling to $K^+K^-$.
The absence of $\phi(1680)$ in $e^+e^-\rightarrow\omega\pi^+\pi^-$ indicates that the $\phi(1680)$ is the promising 2$^3$S$_1$ $s\bar s$ candidate because it prefers to decay into strange mesons as expected from
the OZI rule. The B factories~\cite{bel-bab}
observed the $\phi(1680)$ via the ISR process and found a new decay mode of
$\phi(1680)\rightarrow K^+K^-\pi^+\pi^-$.

The $\phi(1680)$ is also expected to be observed in
photoproduction experiments. However, the latest results on $\gamma
p\rightarrow K^+K^-$ show a clear structure in the
$K^+K^-$ mass spectrum with a mass of $M=1753\pm 3$\,MeV/$c^2$ and a
width of $\Gamma=122\pm63$\,MeV, which is in good agreement with
results from previous photoproduction experiments. The mass is
much higher than the 1680\,MeV/$c^2$ reported from $e^+e^-$ experiments. No structure was observed around the 1.68\,GeV/$c^2$ or
1.75\,GeV/$c^2$ mass regions in the $\gamma p\rightarrow K_S
K^\pm\pi^\mp p$ interaction.
The discrepancy on the measured mass between experiments could be explained by the interference with other mesons. 
The probability that the
structure around 1.75\,GeV/c$^2$ observed in the photoproduction
process is a new state cannot be ruled out~\cite{pdg}.

Therefore, study of the $\phi(1680)$
is necessary to clarify its nature. Based on the
prediction of the $^3P_0$ model, the branching fraction
of $\phi(1680)$ decaying into $\phi\eta$ is smaller than those of
$K\bar{K}$ and $K^*K$. This has not been studied yet, but will play
an important role in addressing the discrepancies.
The data accumulated at BESIII allow us to investigate the
$\phi(1680)$ produced in charmonium decays.

\subsection{$\phi_3(1850)$ ($1^3D_3$)}
The $\phi_3(1850)$ was first reported with a mass of $1850\pm 10$\,MeV/$c^2$
and a width of $80^{+40}_{-30}$\,MeV by a bubble chamber experiment~\cite{hbc} in the $K\bar{K}$ mass spectra in the $K^-p\rightarrow K\bar{K}\Lambda$ interaction. A study of the
$K^*\bar{K}$ mass spectrum in the same experiment also indicated a structure
around 1.85\,GeV/$c^2$ and production rate compatible with the $K\bar{K}$ decay mode. The Omega experiment~\cite{omega} observed a similar structure in the hypercharge exchange reaction $K^-p\rightarrow K\bar{K}(\Lambda/\Sigma^0)$. Its spin-parity was determined to be $J^{PC}=3^{--}$, which is consistent with the absence of $\phi_3(1850)$ in the reaction $K^-p\rightarrow
K^0_S K^0_S\Lambda$. The high statistics data from the LASS
experiment~\cite{lass} confirmed the existence of $\phi_3(1850)$ and the spin-parity of $J^{PC}=3^{--}$. The state is interpreted as the $\phi-$like
of the $J^{PC}=3^{--}$ nonet.

\subsection{$\phi(2170)$ ($3^3S_1$ or $2^3D_1$)}
Another possible strangeonium is the $\phi(2170)$, which was first discovered
with a mass of $2175\pm10\pm15$\,MeV/$c^2$ and
a width of $\Gamma=58\pm16\pm20$\,MeV
in the initial state radiation (ISR) process
$e^+e^-\rightarrow \gamma \phi f_0(980)$~\cite{babar-phi2170}.
The BES experiment confirmed this resonance in the hadronic decay of $J/\psi\rightarrow\phi
f_0(980)\eta$~\cite{bes-phi2170} and the resonance parameters are measured to be $M=2186\pm10\pm6$\,MeV/$c^2$ and $\Gamma=63\pm23\pm17$\,MeV.
In 2007, the Babar experiment updated the results with both decays of
$f_0(980)\rightarrow \pi^+\pi^-$ and $f_0(980)\rightarrow
\pi^0\pi^0$~\cite{babar-phi2170-2}. In addition Babar also observed
an evident structure in the process of $e^+e^-\rightarrow\gamma
\phi\eta$~\cite{babar-phi2170-3}. 
Belle~\cite{belle-phi2170}
also observed this structure. The mass, $2079\pm13^{+79}_{-28}$\,MeV/$c^2$,
is consistent with previous report, while the width $192\pm23^{+25}_{-61}$ MeV, is broader. 
The discovery of $\phi(2170)$ triggered many theoretical speculations on its nature. Ref~\cite{2170-theo} performs a Faddeev calculations for the three mesons system, $\phi K\bar{K}$, and obtain a neat resonance peak around a total mass of 2150\,MeV and an invariant mass for the $K\bar{K}$ system around 970\,MeV. This finding provides a natural explanation for the $\phi(2170)$.
At present, this state is listed in the
PDG~\cite{PDG-2014} as the excited $\phi$ meson, which is also referred to as $Y(2175)$ in the literature. 

The theoretical models~\cite{susanna} present many possible decay modes of the $\phi(2170)$. For most of the dominant decays, e.g., $K^*\bar{K^*}$, $K(1460)\bar{K}$, $K^*(1410)\bar{K}$ and $K_1(1270)\bar{K}$, the mass threshold is quite close to the mass of the $\phi(2170)$, and the broad strange meson also makes it difficult to see a clear structure with the complicated final states. Besides $K\bar{K^*}$, the study
of the decay modes $\phi\eta$ and $\phi\eta^\prime$ are very
helpful to distinguish $\phi(2170)$ among many interpretations.


\begin{table*}[ht]
\begin{center}
\caption{ \label{ssbar} Summary of strangeonium states and their dominant decays.
}

\begin{tabular}{c|c|c|c|c}
\hline
$N^{2S+1}L_{J}$    & state & Mass (MeV/$c^2$)& Width (MeV) & Dominant decays\\
\hline
$1^1S_0$ & $\eta^\prime$ & $(957.78\pm0.06)_{exp.}$    & $(0.198\pm0.009)_{exp.}$  & $\pi\pi\eta$, $\gamma\pi^+\pi^-$\\
$1^3S_1$ & $\phi$        & $(1019.461\pm0.019)_{exp.}$ & $(4.266\pm0.031)_{exp.}$  & $KK$, $\pi^+\pi^-\pi^0$\\
$2^3S_1$ & $\phi(1680)$  & $(1680\pm 20)_{exp.}$       & $(150\pm 50)_{exp.}$      & $K\bar{K}$, $K^*\bar{K}$, $\phi\eta$ \\
$2^1S_0$ & $\eta(1440)$  & $\sim$ 1440                 & $11\sim100$               & $K^*\bar{K}$ \\
$3^3S_1$ & $\phi(2050)$  & $\sim 2050 $                & $\sim 380$                & $K^*\bar{K}$, $K^*\bar{K^*}$, $K_1(1270)\bar{K}$, $K_1(1402)\bar{K}$ , $K^*(1414)\bar{K}$, $\phi\eta$\\
$3^1S_0$ & $\eta_s(1950)$& $\sim 1950 $                & $\sim 175$                &  $K^*\bar{K}$, $K^*\bar{K^*}$, $K^*_{0}(1430)\bar{K}$, $K^*(1414)\bar{K}$\\
\hline

$1^3P_0$ & $f_0(1500)$   & $\sim$ 1500                     & $279$                     & $K\bar{K}$, $\eta\eta$\\
$1^3P_1$ & $f_1(1420)$   & $(1426.4\pm 0.9)_{exp.}$    & $(54.9\pm2.6)_{exp.}$     & $K^*\bar{K}$\\
$1^3P_2$ & $f^\prime_2(1525)$&$(1525\pm5)_{exp.}$      &$(73^{+6}_{-5})_{exp.}$    & $K\bar{K}$, $K^*\bar{K}$, $\eta\eta$ \\
$1^1P_1$ & $h_1(1380)$   & $(1386\pm19)_{exp.}$        &$(91\pm30)_{exp.}$         & $K^*\bar{K}$\\
\hline

$2^1P_1$ & $h_1(1850)$ & $\sim$ 1850  & $193$ & $K^*\bar{K}$, $K^*\bar{K^*}$, $\phi\eta_s$\\
$2^3P_0$ & $f_0(2000)$ & $\sim$ 2000  & $\sim800$ & $K\bar{K}$, $K^*\bar{K^*}$, $K_{1}(1270)\bar{K}$, $\eta\eta_s$\\
$2^3P_1$ & $f_1(1950)$ & $\sim$ 1950  & $\sim300$ & $K^*\bar{K}$ $K^*\bar{K^*}$, $K_{1}(1270)\bar{K}$, $K^*(1410)\bar{K}$\\
$2^3P_2$ & $f_2(2000)$ & $\sim$ 2000  & $\sim400$ & $K^*\bar{K}$, $K^*\bar{K^*}$, $K_{1}(1270)\bar{K}$, $K_2^*(1430)\bar{K}$, $\eta\eta$, $\eta\eta^\prime$\\
\hline

$1^1D_2$ & $\eta_2(1850)$ & $\sim$ 1850  & $129$ & $K^*\bar{K}$, $K^*\bar{K^*}$\\
$1^3D_1$ & $\phi(1850)$   & $\sim$ 1850  & $\sim650$ & $K\bar{K}$, $K^*\bar{K}$, $K_{1}(1270)\bar{K}$,$\phi\eta$\\
$1^3D_2$ & $\phi_2(1850)$ & $\sim$ 1850  & $214$ & $K^*\bar{K}$,$\phi\eta$\\
$1^3D_3$ & $\phi_3(1850)$ & $(1854\pm7)_{exp.}$  & $(87^{+28}_{-23})_{exp.}$ & $K\bar{K}$,$K^*\bar{K}$\\
\hline

$1^1F_3$ & $h_3(2200)$ & $\sim2200$   & $\sim250$  & $K^*\bar{K}$,$K^*\bar{K^*}$, $K^*_2(1430)K$, $K^*K_1(1270)$, $\phi\eta$\\
$1^3F_2$ & $f_2(2200)$ & $\sim$ 2200  & $425$      & $K\bar{K}$,$K^*\bar{K}$,$K^*\bar{K^*}$,$K_1(1270)K$, $K^*_2(1430)K$, $K^*K_1(1270)$\\
$1^3F_3$ & $f_3(2200)$ & $\sim$ 2200  & $\sim300$  & $K^*\bar{K}$,$K^*\bar{K^*}$,$K_1(1270)K, K^*_2(1430)K$, $K^*K_1(1270)$\\
$1^3F_4$ & $f_4(2200)$ & $\sim$ 2200  & $\sim150$  & $K\bar{K}$ $K^*\bar{K}$,$K^*\bar{K^*}$\\
\hline
\end{tabular}
\end{center}
\end{table*}

\subsection{Other strangeonium states}
Based on the expectations of the $^3P_0$ model, orbital and radial excited strangeonium states
are summarized in
Table~\ref{ssbar}, where the masses and widths are from the world average values in Ref.~\cite{pdg} or Ref.~\cite{p3model}. Most of the unobserved
strangeonium states are in the $1\sim2$\,GeV/$c^2$ mass range; 
their widths are expected to be broad. Due to the overlap with the
$q\bar{q}$ ($q=u,d$) states, it is hard for an experiment to observe them
by bump hunting alone. Thus, high statistics experiments with large acceptance
spectrometers and full PWA in several different channels are required
to sort out the multi-states and to determine their quantum numbers.
The application of the PWA technique necessarily includes information about
normal $q\bar{q}$ mesons, both established states and undiscovered. This also applies to the identification of the hybrid mesons
with non-exotic quantum numbers that can mix with normal $q\bar{q}$ states.
Thus, as part of the program of identifying hybrid mesons, the high
statistics data sets will enable the study of ground state $q\bar{q}$
mesons as well.

\subsection{Current experiments studying strangeonia}
\begin{table*}[ht]
\begin{center}
\caption{
\label{sum_exp} Summary of current or future experiments studying
strangeonium physics.}
\begin{tabular}{c|c|c}
\hline
Experiments  & Timeline & Production\\
\hline
BESIII & 2008 & Charmonia decays, $e^+e^-$ annihilations \\
CMD-3 & 2010 & $e^+e^-$ annihilations\\
GlueX & 2015 & photoproduction\\
CLAS12 & 2015 & photoproduction \\
BelleII & 2016 & ISR\\
PANDA &2018? & hadronic production\\
\hline
\end{tabular}
\end{center}
\end{table*}

Different experiments, using $e^+e^-$ collision, photoproduction ($\gamma p$), and hadron production ($K p$), have contributed greatly to
the understanding of the $s\bar{s}$ spectrum. However, as discussed above,
our present information on the strangeonium spectrum is still far from
complete. Even for the established strangeonium states, there are still many unresolved issues. Due to the low production rates, broad widths and
complex final states, an experiment must have both much higher statistical sensitivity and good acceptance to make a significant contribution to the strangeonium sector. Table~\ref{sum_exp} summarizes current and future experiments which have the capability of exploring the strangeonium states.

The BESIII detector~\cite{bes3} fulfils these requirements with its excellent
performance. The superconduction solenoid and the
helium-based drift chamber, covering 93\% of 4$\pi$ stereo angle,
give high acceptance and good momentum resolution; the combination
of the Time-Of-Flight (TOF) and d$E$/dx measurements provide good
particle identification; the high-performance CsI calorimeter has an
energy resolution of 2\% for 1\,GeV photons. The available high
statistics data offer a unique opportunity to
study strangeonium spectroscopy.

In the case of photoproduction, Jefferson Lab~\cite{jlab} is completing its
upgrade of the CEBAF accelerator to 12\,GeV electron beam, along
with new installations to study meson spectroscopy. The GlueX detector in Hall-D is designed to have a uniform acceptance overall and cover all decay angles,
which is essential for amplitude analysis. Using the linearly
polarized photon produced by the coherent Bremsstrahlung of the
primary electron beam on a diamond radiator, GlueX is capable of
searching for exotics as well as performing conventional meson spectroscopy.

With the Forward Tagger Facility, CLAS12 in Hall-B, which has been designed to determine the Generalized Parton Distributions (GPDs), will also have the capability to study the meson
spectrum using virtual photons.

In the last decade, the KLOE experiment at DA$\Phi$NE played an important role in the study
of $\phi$, $\eta$ and $\eta^\prime$ decays. The KLOE-2 experiment has a plan to collect about 50\,fb$^{-1}$ in several years, which will investigate $\phi$, $\eta$ and $\eta^\prime$ decays with unprecedented
precision. The other two $e^+e^-$ experiments, CMD-3 and
SND at VEPP-2000, are collecting data at the center-of-mass of
$0.3\sim2.0$\,GeV, which mainly focus on the physics in the low energy region.
Recently a preliminary result on the observation of $\phi(1680)$
was reported by CMD-3. The alternative strategy is to use
B-factory data, after radiation of a hard initial state photon,
to investigate the light hadron spectroscopy; many nice results from Belle and Babar
have been summarized in Ref.~\cite{isr-rev}.
In the near future, the Belle-II detector will collect an enormous amount
of data for the study of heavy flavor physics, as well as the study of the strangeonium sector with high precision.

On hadron production, the PANDA experiment being constructed at FAIR
is a general purpose spectrometer
that will map the hadron spectrum,
using antiproton beams colliding with an
internal proton target. Based on simulations and
the studies of specific physics channels, PANDA will be excellent~\cite{09033905}
at distinguishing states of interest from the huge number of background
events.

\section{Study of $s\bar{s}$ states through charmonium decays}

According to the predictions from the $^{3}P_{0}$ model,
most of these states are broad and the decay final states
are complicated, so they are usually indistinguishable in the mass spectrum.
We therefore need to have an experiment with
excellent charged and neutral detection and high statistics to hunt for those states from a large background.
The BESIII detector at the BEPCII, running at the energy range of 2$-$4.6\,GeV, may provide an
important contribution to this field. The BESIII detector is a large
solid-angle magnetic spectrometer with high acceptance, full primary
vertex reconstruction and secondary vertex reconstruction for
long lived particle such as $K^0_{S}$ and $\Lambda$, high
particle identification efficiency, good momentum resolution of
charged particles and excellent energy resolution of electrons and
photons. In addition, the designed peak luminosity of BEPCII,
$10^{33}$ cm$^{-2}s^{-1}$ at 3.773\,GeV, is about 100 times better
than its predecessor, which allows us to accumulate the very large
data samples in a short period of time. About 1.3 billion
$J/\psi$ events\cite{jpsi} and 0.5 billion $\psi(2S)$ events have been collected by
the BESIII detector, which will provide a great opportunity to perform
experimental studies of strangeonium produced in $J/\psi$ and
$\psi(2S)$ decays.

The world's largest direct data $J/\psi$, $\psi(2S)$ samples, collected by the BESIII
experiment, offer excellent opportunities to detect new $s\bar{s}$ mesons and new states
through the decays of the $J/\psi$ and $\psi(2S)$. Many $J/\psi$, $\psi(2S)$ decays to $\phi$
and $\eta$ have been well measured. Examples of such decays are $J/\psi\rightarrow \eta K^*\bar{K}^*, \phi K^*\bar{K}, \phi K\bar{K}, \phi f, \phi\pi\pi, \phi\eta, \eta\pi\pi$.
Possible new $s\bar{s}$ states can be probed through the recoil masses of the $K^*\bar{K}^*, K^*\bar{K}, K\bar{K}, f, \pi\pi, \eta / \phi$, respectively, based on the expectation that the same processes that produce the $\phi, \eta$ in the $J/\psi$, $\psi(2S)$ decays will couple to new $s\bar{s}$ mesons which are kinematically allowed.

The $J/\psi\rightarrow\gamma\eta^\prime$ is a two-body decay and the
photon energy is monochromatic, which makes it easy to
distinguish the $\eta^\prime$ from the background. Considering
the large branching fraction of $J/\psi\rightarrow\gamma\eta^\prime$, the $J/\psi$ data sample offers a clean
environment to investigate $\eta^\prime$ anomalous decays.

The $\phi$ production rate in $J/\psi$ hadronic decays is at a level
of $3 \times 10^{-4} \sim 2 \times 10^{-3}$.
The $\phi$ could be easily reconstructed with its dominant $K^+K^-$
decay mode.
Of interest is to investigate the mass spectrum
recoiling against $\phi$ to search for the rare decays of
$\eta/\eta^\prime$ via the two-body decays of
$J/\psi\rightarrow\phi\eta/\eta^\prime$~\cite{raredecay}. In
addition, the precision measurement of the full set of $J/\psi$
decays into a vector and a pseudoscalar pair are also allowed to
investigate the pseudoscalar mixing and the gluonic content of the
$\eta^\prime$.

According to the $^3P_0$ model, the $h_1(1380)$ dominantly decays into $K^*K$, whereas $\omega\eta$, and $\rho\pi$ are
suppressed due to the OZI rule~\cite{p3model2}. The $\phi\eta$ mode could be its favorable decay
mode, but is strongly suppressed because of the limited phase space. Therefore the $J/\psi\rightarrow K^*K\eta$ and $J/\psi\rightarrow K^*K\eta^\prime$ modes would be the most preferable channels to study the $h_1(1380)$ at BESIII. In addition, the radiative decays of $h_1(1380)$ to $\eta$ or $\eta^\prime$ may be reconstructed in $J/\psi$ data, though there is no theoretical prediction for their branching ratios.

\section{Hunt for strangeonium-like particles at BESIII}
In 2013, BESIII and Belle
observed a charged charmonium-like $Z_c^+(3900)$~\cite{zc3900} state, and subsequently several similar structures were reported by the BESIII experiment. These observations inspired an extensive discussion on their internal substructures.
Most recently, a neutral partner of the $Z^{\pm}_c(3900)$ has been observed, which indicates that $Z^+_c(3900)$ is part of an isotriplet and suggests a new hadron spectroscopy.

We propose a search for the charged strangeonium-like structure
in the decay $\phi(2170)\rightarrow \pi^+\pi^-\phi$, as in Ref.~\cite{xliu}. Similar to $Y(4260)\rightarrow \pi^+\pi^- J/\psi$ and
$\Upsilon(10860)\rightarrow\pi^+\pi^-\Upsilon(1S,2S)$, two charged
strangeonium-like strucutures are expected to be observed in
$\phi(2170)\rightarrow \pi^+\pi^-\phi$.
Therefore, $\phi\pi$ is an ideal channel to detect
strangeonium-like states. Since the isospin of the $\phi\pi$
system is 1, the isosinglet $s\bar{s}$ state decaying into $\phi\pi$ is suppressed by isospin violation. 
For the conventional mesons composed of
$u,d$ quarks, the $\phi\pi$ decay mode is strongly suppressed by
the OZI rule. Therefore the observation of a $\phi\pi$ decay mode
may imply an exotic nature. 

At BESIII, different processes, including ISR, $J/\psi$ decays and data
taken at the peak of the $\phi(2170)$ could be used to make an extensive study of the
$\phi(2170)$.
First, the ISR process $e^+e^-\rightarrow \gamma_{ISR}(\phi')$ could also be used to study $\phi(2170)\rightarrow\phi
\pi\pi$. The most significant data samples recorded by \mbox{BESIII} are \mbox{$\sim2.9$\,fb$^{-1}$} $\psi(3770)$ and $\sim3$\,fb$^{-1}$
above 3.8\,GeV. They are not sufficient for an extensive study of the
$\phi(2170)$.

For the $\phi$(2170) in $J/\psi\rightarrow\eta \phi\pi^+\pi^-$, the background level is quite high under the $\phi(2170)$,
which makes it hard to investigate the $\phi\pi$ for any states with small
production rate. Comparing to the other $e^+e^-$ experiments, \mbox{BEPCII} has an
advantage in taking data directly at 2.2\,GeV. 
With the assumption of 100\,pb$^{-1}$ data at the peak of $\phi$(2170) and $\mathcal{B}(\phi(2170)\rightarrow Z_s\pi)\sim$ 10\% $\times \mathcal{B}(\phi(2170)\rightarrow \phi f_0(980))$, the observed number of $Z_s$ events is estimated to be
$L\times 10\% \times \sigma(\phi(2170)\rightarrow \phi
f_0(980))\times \varepsilon\times \mathcal{B}(\phi\rightarrow K^+K^-) \sim 300$, where the efficiency is estimated to be about 50\%.
This may enable us to
investigate the existence of $Z_s$.

\section{Search for new $s\bar{s}$ mesons in $e^+e^-\rightarrow (s\bar{s})(s\bar{s}$) interactions}
We propose to search for new $(s\bar{s})$ mesons produced in association with a well established $(s\bar{s})$ state such as the $\phi$ or $\eta/\eta'$.
This is motivated by the large $(c\bar{c})(c\bar{c})$ rates reported by the Belle~\cite{Belle-jpsiccbar} and BaBar~\cite{BABAR-jpsiccbar} experiments which have helped establish the $\eta_{c}(2S)$ state in the recoil mass spectra against a $J/\psi$ or a $\psi(2S)$.

Though the mechanism through which large $e^+e^- \rightarrow J/\psi, \psi(2S)+c\bar{c}$ interactions occur is not well understood, if the same process works for the $e^+e^- \rightarrow (s\bar{s})(s\bar{s})$ process, we can probe a $(s\bar{s})$ system with a fully reconstructed $\phi$, $\eta$ or $\eta'$ with the BESIII data by examining the recoil side of the event, for which the recoil mass can be calculated by $M_{recoil} = \sqrt{(\sqrt{s}-E_{\phi}^{*})^{2} - p_{\phi}^{* 2}}$. Here the center of mass energy $\sqrt{s}$ is well measured, 
and the $\phi$, $\eta$ and $\eta'$ mesons are all narrow, resulting a very good resolution on the $M_{recoil}$. This approach can probe the $(s\bar{s})$ system and detect $(s\bar{s})$ states on the recoil side without the need to reconstruct specific final states of the $(s\bar{s})$ meson in the recoil.

\section{Summary}
Though 
a substantial number of hadrons have been established experimentally,
there are predicted light hadrons in the mass region of 1 to 2\,GeV/$c^2$, which
have not been observed yet. Many of the missing hadrons are
strangeonium states.
Techniques for searching for new $s\bar{s}$ mesons with the BESIII detector, and topics related to the $s\bar{s}$ states, have been outlined in this paper.

High statistics data accumulated with the \mbox{BESIII} detector
are important for the investigation of the strangeonium spectroscopy, and will help
distinguish exotic states from conventional mesons. BESIII data can
enable probing of the meson spectrum with unprecedented detail.

\end{multicols}

\clearpage
\end{document}